 \documentclass[prd,preprintnumbers,amsmath,amssymb,floatfix]{revtex4}

\usepackage{graphicx}
\usepackage{dcolumn}
\usepackage{bm}
\usepackage{amssymb}
\usepackage{epsfig}
\usepackage{color}

\newcommand{\ba}{\begin{eqnarray}}
\newcommand{\ea}{\end{eqnarray}}
\newcommand{\be}{\begin{equation}}
\newcommand{\ee}{\end{equation}}
\newcommand{\bdisplay}{\begin{displaymath}}
\newcommand{\edisplay}{\end{displaymath}}

\newcommand{\eq}[1]{Eq.\,(\ref{#1})}

\newcommand{\chii}{\chi_{{}_{\rm I}}}

\newcommand{\sigtot}{\sigma_{\rm tot}}

\begin{document}

\title{ Forward hadronic scattering at 7 TeV: predictions for the LHC; an update.}  
\author{Martin~M.~Block}
\affiliation{Department of Physics and Astronomy, Northwestern University, 
Evanston, IL 60208}
\author{Francis Halzen}
\affiliation{Department of Physics, University of Wisconsin, Madison, WI 53706}
\date{\today}

\begin{abstract}
The LHC has successfully run for a long period at half energy, 7 TeV. 
In this note, we update earlier full-energy  Large Hadron Collider (LHC) forward hadronic scattering predictions \cite{physicsreports}, giving new  predictions, including errors, for the  $pp$ total and inelastic cross sections, the  $\rho$-value,  the  nuclear slope parameter $B$, $d\sigma_{\rm el}/dt$, and the large gap survival probability  at the current 7 TeV energy.
 \end{abstract}

\maketitle


\section{Introduction} \label{sec:introduction} 
The LHC has run at 7 TeV (half-energy) for an extended period of time and large amounts of data have been collected. Five years ago, we made hadronic forward scattering predictions for the full energy (14 Tev) Large Hadron Collider; for details see the review article by M. Block  \cite{physicsreports}. Recently, we have had inquiries from LHC experimental groups for 7 TeV predictions.  The purpose of this note is to gather together in one convenient location  an update to the 2006 publication, in which we furnish comparisons between new 7 TeV and (already published) 14 TeV results, including errors in the 7 TeV predictions due to model uncertainties, as well as presenting new calculations for $pp$ elastic scattering, $d\sigma/dt$, at 7 TeV. We have combined two separate models to make these predictions, the first being the analyticity-constrained analytic amplitude model of Block and Halzen \cite{newfroissart} that saturates the Froissart bound \cite{froissart} and the second being the ``Aspen Model'', a revised  version of the eikonal model of Block, Gregores, Halzen and Pancheri \cite{aspenmodel} that now incorporates analyticity constraints. We  purposely keep explanations very brief; for complete details, see Ref. \cite{physicsreports}.

\section{Predictions at 7 TeV } \label{sec:predictions}

\subsection{The analytic amplitude model}
\begin{figure}[h,t,b] 
\begin{center}
\mbox{\epsfig{file=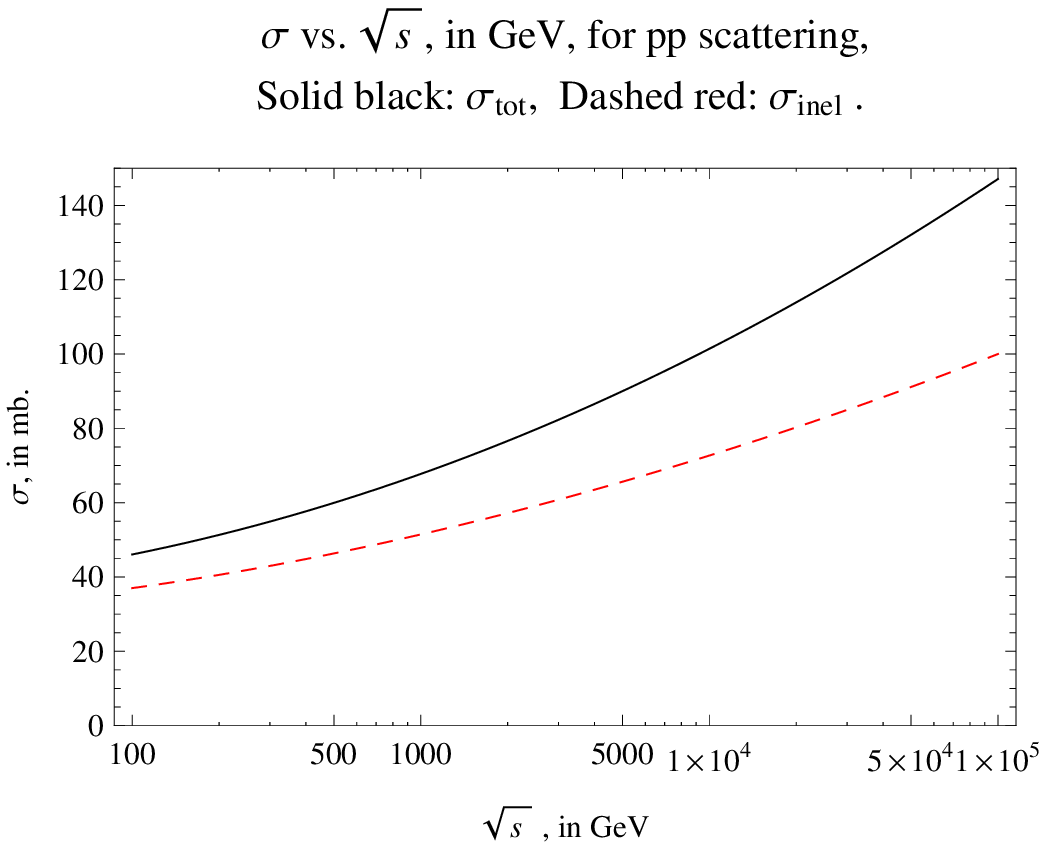
,width=3in%
,bbllx=0pt,bblly=0pt,bburx=300pt,bbury=198pt,clip=%
}}
\end{center}
\caption[]{
The total $pp$ cross section, $\sigma_{\rm tot}$, and the inelastic cross section  $\sigma_{\rm inel}$, in mb, vs. $\sqrt s$, the cms energy in GeV. The solid (black) curve is the total cross section and the dashed (red) curve is the inelastic cross section.\label{fig:sigtot}
}
\end{figure}

We make the most accurate predictions of  the forward $pp$ scattering properties,
\ba
\sigma_{\rm tot}&\equiv&{4\pi\over p} {\rm Im}f(\theta_L=0)\label{sigtot1}\\
\rho&\equiv&{{\rm Re}f(\theta_L=0 )\over {\rm Im}f(\theta_L=0)},\label{rho}
\ea
using the analyticity-constrained analytic amplitude model of Block and Halzen \cite{newfroissart} that saturates the Froissart bound \cite{froissart}. By saturation of the Froissart bound, we mean  that the total cross section defined in \eq{sigtot1} rises as $\ln^2 s$, where $s$ is the square of the cms energy. In \eq{sigtot1} and \eq{rho}, $f(\theta_L)$ is the $pp$  laboratory scattering amplitude as a function of $\theta_L$, the laboratory scattering angle and $p$ is the laboratory momentum. In Fig. \ref{fig:sigtot}, the solid line is  the total pp cross section as a function of the cms energy, $\sqrt s$. Our use of analyticity constraints---employing new Finite Energy Sum Rules (FESR) \cite {FESR}---allows us to use {\em very accurate  low energy cross section measurements to act as an anchor} that accurately fixes our high energy cross section predictions.  

At 7 TeV, we find that $\sigma_{\rm tot}=95.4\pm 1.1$ mb. The same set of parameters  predict $\sigma_{\rm tot}=107.3\pm 1.2$ mb at 14 TeV \cite{physicsreports}.  Further,  at 7 TeV we predict that $\rho_{pp}=0.135\pm 0.001$. 
\subsection{The``Aspen'' Model: an eikonal model for $pp$ scattering}
The ``Aspen'' model uses an unconventional  definition of the  eikonal $\chi(b,s)$ in impact parameter space $b$, so that 
\ba
\sigtot(s)&=&2\int \left[1-e^{-\chi_I(b,s)}\cos\left(\chi_R(b,s)\right)\right]\, d^2\vec b,\label{sigelofb20}\\
\rho(s)&=&\frac{\int e^{-\chi_I(b,s)}\sin(\chi_R(b,s))\,d^2\vec b}{\int\left[ 1-e^{-\chi_I(b,s)}\cos(\chi_R(b,s))\right]\,d^2\vec b}\quad, \label{rhoofb0}\\
B(s)&=&\frac{1}{2}\frac{\int | e^{-\chi_I(b,s)+i\chi_R(b,s)}-1|b^2\,d^2\vec b}{\int | e^{-\chi_I(b,s)+i\chi_R(b,s)}-1|\,d^2\vec b},\label{B}\\
\frac{d\sigma_{\rm el}}{dt}&=&\pi\left|\int J_0(qb)\left[ e^{-\chi_I(b,s)+i\chi_R(b,s)}-1\right]b\,db\right|^2,\label{dsdt}\\
\sigma_{\rm el}(s)&=&\int\left| e^{-\chi_I(b,s)+i\chi_R(b,s)}-1\right|^2\,d^2\vec b,\\
\sigma_{\rm inel}(s)&\equiv&\sigtot(s) -\sigma_{\rm el}(s)=\int \left( 1-e^{-2\chi_I(b,s)}\right)\,d^2\vec b,\label{siginel}
\ea 
where $\sigma_{\rm inel}(s)$ is the total inelastic cross section.
The even eikonal profile function $\chi^{ even}$, which is the only surviving term at the high energies considered here,  receives contributions 
from
quark-quark, quark-gluon and gluon-gluon interactions, and can be written in the factorized form
\begin{eqnarray}
\chi^{ even}(s,b) &=& \chi_{qq}(s,b)+\chi_{qg}(s,b)+\chi_{gg}(s,b)
\nonumber \\
&=& i\left [ \sigma_{qq}(s)W(b;\mu_{qq})
+ \sigma_{qg}(s)W(b;\sqrt{\mu_{qq}\mu_{gg}})
+ \sigma_{gg}(s)W(b;\mu_{gg})\right ]\, ,\label{chiintro}
\end{eqnarray}
where $\sigma_{ij}$ is the cross sections of the colliding partons, and
$W(b;\mu)$ is the overlap function in impact parameter space,
parameterized as the Fourier transform of a dipole form factor. The parameters $\mu_{qq}$ and $\mu_{gg}$ are masses which describe the ``area'' occupied by the quarks and gluons, respectively, in the colliding protons. In this model hadrons asymptotically evolve into black disks of
partons. For details of the parameterization of the model, see Ref. \cite{physicsreports}.

From \eq{siginel} and \eq{sigelofb20}, we calculate the {\em ratio} $r(s)=\sigma_{\rm inel}(s)/\sigma_{\rm tot}(s)$, because most errors due to  parameter uncertainties cancel in the ratio.  We then multiply $r(s)$ by the (more accurate) total cross section using \eq{sigtot1} (the analytic amplitude model) to obtain the inelastic cross section shown in Fig. 
\ref{fig:sigtot}, as the dashed (red) curve.  At 7 TeV, we find $\sigma_{\rm inel}=69.0\pm 1.3$ mb.

Further, from, \eq{B} we find that the nuclear slope parameter $B$, the logarithmic derivative of the elastic cross section (as a function of squared momentum transfer $t$) with respect to $t$, at $t=0$ is given by $B= 18.28\pm 0.12$ (GeV/c)$^{-2}$ at 7 TeV.   

\begin{figure}[h,t,b] 
\begin{center}
\mbox{\epsfig{file=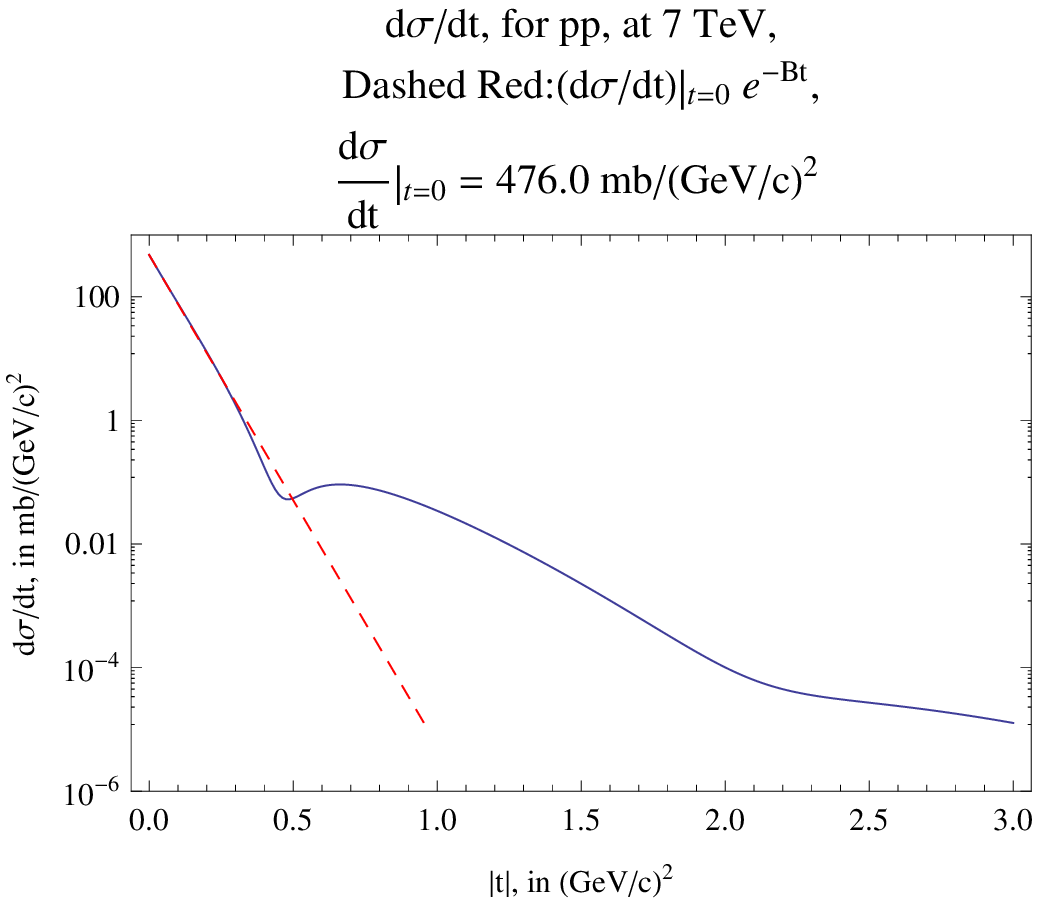
,width=3in%
,bbllx=0pt,bblly=0pt,bburx=300pt,bbury=192pt,clip=%
}}
\end{center}
\caption[]{
The $pp$ differential elastic scattering cross section  , $d\sigma_{\rm el}/dt$, in mb/(GeV/c)$^2$, vs. $|t|$, in (GeV/c)$^2$ is the solid (black) curve. The dashed (red) curve is the small $|t|$ approximation, ${d\sigma \over dt}|_{t=0} e^{-B|t|}$.\label{fig:dsdt}
}
\end{figure}

At 7 TeV, using \eq{dsdt},  we plot the differential elastic scattering cross section $d\sigma_{\rm el}/dt$, in mb/(GeV/c)$^2$, against $|t|$, in (Gev/c)$^2$ as the solid (black) curve in Fig. \ref{fig:dsdt}.  Also shown is the approximation,  ${d\sigma \over dt}|_{t=0} e^{-B|t|}$,  valid for small $|t|$, which is the dashed (red) curve. The agreement is striking for small $t$. 
\subsection{Rapidity gap survival probabilities}
\begin{figure}[h,t,b] 
\begin{center}
\mbox{\epsfig{file=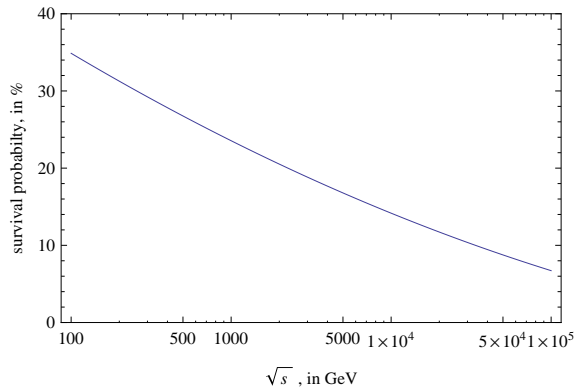
,width=3in%
,bbllx=0pt,bblly=0pt,bburx=300pt,bbury=200pt,clip=%
}}
\end{center}
\caption[]{
 $<|S|^2>$, the survival probability of large rapidity gaps in $pp$ collisions, in \%, vs. $\sqrt s$, the cms energy in GeV.\label{fig:survival}
}
\end{figure}

As shown in Ref. \cite{physicsreports}, the survival probability $<|S|>^2$  of {\em any} large rapidity gap is given by
\be 
<|S|^2>=\int W(b\,;\mu_{\rm qq})\,e^{-2\chii(s,b)}d^2\,\vec{b},\label{eq:survival}
\ee 
which is the differential probability density in impact parameter space $b$ for {\em no} subsequent interaction (the exponential suppression factor)  multiplied by the quark probability distribution in $b$ space from \eq{chiintro}), which is then integrated over $b$.  It should be emphasized that \eq{eq:survival} is the probability of {\em survival} of a large rapidity gap and {\em not} the probability for the production and survival of large rapidity gaps, which is the quantity observed experimentally. The energy dependence of the survival probability $<|S|^2>$ is through the energy dependence of $\chii$, the imaginary portion of the eikonal given in \eq{chiintro}. A plot of $<|S|^2>$ as a function of $\sqrt s$, the cms energy in GeV, is given in Fig. \ref{fig:survival}. At 7 TeV, we find the gap survival probability to be  $<|S|^2>=15.5\pm 0.05$ \%.
\section{Summary}We summarize our $pp$ forward scattering parameters for the LHC in Table \ref{table:LHC}, comparing the 7 and 14 TeV values. The 14 TeV values are taken from Ref.\cite{physicsreports}.

\begin{table}[h,t]                   
%
\def\arraystretch{1.5}            

\begin{center}
\caption[]{Values of forward scattering parameters for the LHC, at 7 and 14 TeV.
\label{table:LHC}
}
\vspace{.2in}
\begin{tabular}[b]{|c||c|c|c|c|c|}
\hline\hline
$\sqrt s$&$\sigma_{\rm tot}$&$\sigma_{\rm inel}$&$\rho$&$B$& $<|S|^2>$\\
(TeV)&mb&mb&&(GeV/c)$^{-2}$&\%\\
\hline
7&$95.4\pm 1.1$&$69.0\pm1.3$&$0.135\pm 0.001$&$18.28\pm 0.12$&$15.5\pm 0.05$\\
\hline
14&$107.3\pm 1.2$&$76.3\pm1.4$&$0.132\pm 0.001$&$19.39\pm0.13$&$12.6\pm0.06$\\
\hline\hline
\end{tabular}
\end{center}
\end{table}
\def\arraystretch{1}  
\begin{acknowledgments}
The work of F.H. is supported in part by the National Science Foundation under Grant No. OPP-0236449, by the DOE under grant DE-FG02-95ER40896 and in part by the University of Wisconsin Alumni Research Foundation.
One of us (M.M.B.) would like to thank  the Aspen Center for Physics for its hospitality during the time parts of this work were done.  
\end{acknowledgments}


\end{document}